\input{aipcheck}
\documentclass[sort&compress,final]{aipproc}

\layoutstyle{6x9}
\usepackage{epsfig}
\usepackage{amsmath}
\usepackage{amssymb}
\usepackage{array}

\newcommand{\crit}[1]{\ensuremath{{#1}^{\star}}}

\newcommand{\sat}[1]{{#1}_\mathnormal{\sigma}}
\newcommand{\liq}[1]{{#1}'}
\newcommand{\qq}[1]{\lq{#1}\rq}
\newcommand{\vap}[1]{{#1}''}

\newcommand{\absnumof}[1]{\ensuremath{{\absnum}_{#1}}}
\newcommand{\areaof}[1]{\ensuremath{{\area}({#1})}}
\newcommand{\areasphof}[1]{\ensuremath{{\area_\bullet}({#1})}}
\newcommand{\chempotof}[1]{\ensuremath{{\chempot}_\mathrm{\ell}}}
\newcommand{\Deltagrandpotentialof}[1]{\ensuremath{\Delta{\grandpotential}_{#1}}}
\newcommand{\nuclrateof}[1]{\ensuremath{{\nuclrate}_{#1}}}
\newcommand{\probgrowof}[1]{\ensuremath{\mathnormal{\probgrow({#1})}}}
\newcommand{\probshrinkof}[1]{\ensuremath{\mathnormal{\probshrink({#1})}}}

\newcommand{\probinftyof}[1]{\ensuremath{\probinfty_{#1}}}
\newcommand{\probfinallyof}[1]{\ensuremath{\mathnormal{\probfinally({#1})}}}

\newcommand{\absnum}{\ensuremath{\mathnormal{N}}}
\newcommand{\area}{\ensuremath{\mathcal{F}}}
\newcommand{\chempot}{\ensuremath{\mathnormal{\mu}}}
\newcommand{\chempotid}{\ensuremath{\chempot_\mathrm{id}}}

\newcommand{\curvedtension}{\ensuremath{\tilde{\tension}}}
\newcommand{\delay}{\ensuremath{\tau}}
\newcommand{\density}{\ensuremath{\mathnormal{\rho}}}
\newcommand{\differential}{\ensuremath{\mathnormal{d}}}

\newcommand{\grandpotential}{\ensuremath{\mathnormal{\Omega}}}

\newcommand{\invkT}{\ensuremath{\mathnormal{\beta}}}
\newcommand{\kboltz}{\ensuremath{\mathnormal{k}_\mathrm{B}}}
\newcommand{\landau}{\ensuremath{\mathcal{O}}}
\newcommand{\liqdensity}{\ensuremath{\density_\mathrm{\ell}}}
\newcommand{\LFKa}{\ensuremath{\alpha_1}}
\newcommand{\LFKb}{\ensuremath{\alpha_2}}
\newcommand{\LJeps}{\ensuremath{\mathnormal{\varepsilon}}}
\newcommand{\LJsig}{\ensuremath{\mathnormal{\sigma}}}
\newcommand{\mass}{\ensuremath{\mathnormal{m}}}
\newcommand{\noniso}{\ensuremath{\mathnormal{f_{\Delta\temperature}}}}
\newcommand{\nucdensity}{\ensuremath{\density_\mathrm{n}}}
\newcommand{\nucpressure}{\ensuremath{\pressure_\mathrm{\ell}}}
\newcommand{\nuclsize}{\ensuremath{\mathnormal{\nu}}}
\newcommand{\nuclrate}{\ensuremath{\mathcal{J}}}
\newcommand{\nuclratesim}{\ensuremath{\nuclrate}}
\newcommand{\nuclrateCNT}{\ensuremath{\nuclrate_\mathrm{CNT}}}
\newcommand{\nuclrateLFK}{\ensuremath{\nuclrate_\mathrm{LFK}}}
\newcommand{\nuclrateHale}{\ensuremath{\nuclrate_\mathrm{HSL}}}

\newcommand{\nucvolume}{\ensuremath{\volume_\mathrm{\ell}}}

\newcommand{\partition}{\ensuremath{\mathnormal{W}}}
\newcommand{\pconstant}{\ensuremath{\mathnormal{\digamma}}}
\newcommand{\planartension}{\tension}
\newcommand{\planck}{\ensuremath{\mathnormal{h}}}

\newcommand{\pressure}{\ensuremath{\mathnormal{p}}}
\newcommand{\probability}{\ensuremath{\mathcal{P}}}
\newcommand{\probfinally}{\ensuremath{\probability^\mathsf{F}}}
\newcommand{\probgrow}{\ensuremath{\probability^+}}
\newcommand{\probshrink}{\ensuremath{\probability^-}}
\newcommand{\probinfty}{\ensuremath{q}}

\newcommand{\supersat}{\ensuremath{\mathnormal{S}}}
\newcommand{\surfaceenergy}{\ensuremath{\mathnormal{\eta}}}
\newcommand{\temperature}{\ensuremath{\mathnormal{T}}}

\newcommand{\Ttr}{\ensuremath{\temperature_3}}
\newcommand{\tension}{\ensuremath{\mathnormal{\gamma}}}
\newcommand{\thermal}{\ensuremath{\mathnormal{\Lambda}}}
\newcommand{\threshold}{\ensuremath{\mathnormal{\Theta}}}
\newcommand{\timea}{\ensuremath{\mathnormal{t}}}

\newcommand{\upot}{\ensuremath{\mathcal{V}}}
\newcommand{\volume}{\ensuremath{\mathnormal{V}}}
\newcommand{\zeldovich}{\ensuremath{\mathnormal{f}_\mathrm{Z}}}

\newcommand{\daemon}{d\ae{}mon}
\newcommand{\LJTS}{t.\ s.\ LJ}

\newcommand{\nucleus}{cluster}
\newcommand{\nuclei}{clusters}
\newcommand{\Vapour}{Vapour}
\newcommand{\vapour}{vapour}
\newcommand{\vapours}{vapours}

\begin{document}

\title{Steady-state molecular dynamics simulation of vapour to liquid nucleation with McDonald's d\ae{}mon}


\author{Martin Horsch}{
  address={Universit\"at Paderborn, Institut f\"ur Verfahrenstechnik,
                                    Warburger Str.\ 100, 33098 Paderborn, Germany}
}
\author{Svetlana Miroshnichenko}{
  address={Universit\"at Paderborn, Institut f\"ur Verfahrenstechnik,
                                    Warburger Str.\ 100, 33098 Paderborn, Germany}
}
\author{Jadran Vrabec\footnote{
      Author to whom correspondence should be addressed: Prof.\ Dr.-Ing.\ habil.\ J.\ Vrabec. E-mail: \underline{jadran.vrabec@upb.de}.
   }
  }{
  address={Universit\"at Paderborn, Institut f\"ur Verfahrenstechnik,
                                    Warburger Str.\ 100, 33098 Paderborn, Germany}
}

\begin{abstract}
The most interesting step of condensation is the cluster
formation up to the critical size. In a closed system, this is an instationary
process, as the vapour is depleted by the emerging liquid phase.
This imposes a limitation on direct molecular dynamics (MD) simulation of nucleation
by affecting the properties of the vapour to a significant extent so that
the nucleation rate varies over simulation time.
Grand canonical MD with McDonald's d\ae{}mon is discussed in the 
present contribution and applied for sampling both nucleation
kinetics and steady-state properties of a supersaturated vapour.

The idea behind that approach is to simulate the production of clusters
up to a given size for a specified supersaturation. 
In that way, nucleation is studied by a steady-state simulation.
A series of simulations 
is conducted for the truncated and shifted Lennard-Jones fluid
which accurately describes the fluid phase coexistence of noble gases and methane.
The classical nucleation theory is found to overestimate the
free energy of cluster formation and to deviate by two orders of magnitude
from the nucleation rate below the triple point at high supersaturations.
\end{abstract}

\maketitle

\section{Introduction}
The key properties of nucleation processes in a super\-saturated \vapours{} are the
height $\crit{\Delta\grandpotential}$ of the free energy barrier that must be
overcome to form stable \nuclei{} and the nucleation rate $\nuclrate$ that indicates
how many macroscopic droplets emerge in a given volume per time.
The most widespread approach for calculating these quantities is
the classical nucleation theory (CNT) \cite{FRLP66}, which has significant
shortcomings, e.g., it overestimates the free energy of \nucleus{}
formation \cite{Talanquer07, CKKN08}.
An important problem of CNT in case of \vapour{} to liquid nucleation
is that the underlying basic assumptions for the liquid do not apply
to nanoscopic \nuclei{} \cite{Tolman49, Bartell01, KW02}.

Molecular simulation permits the investigation of nanoscopic surface
effects and the stability of supersaturated states from first
principles, using effective pair potentials. For instance, the
spinodal line can be detected with Monte Carlo (MC) \cite{SE04}
simulation methods;
in experiments, it can only be approximated
as it is impossible to discriminate an unstable state from a metastable
state where $\crit{\Delta\grandpotential}$ is low.
Equilibria \cite{VKFH06} and vapourization processes \cite{IHHK96, SF05a}
of single \nuclei{} can also be simulated to obtain the surface tension as
well as heat and mass transfer properties of strongly curved interfaces.
Moreover, molecular dynamics (MD) \cite{YM98, RK07b, HVH08} 
and MC \cite{NV05} simulation of supersaturated systems with a large number of particles are useful
for the study of very fast nucleation processes, whereas lower nucleation rates can
be calculated by transition path sampling based methods \cite{VAMFT07, MPSF08}.

Equilibrium simulations fail
to reproduce kinetic properties of nucleation processes
such as the overheating of growing \nuclei{} due to latent heat.
On the other hand, direct MD simulation of
nucleation, where \nucleus{} formation is observed directly in a near-spinodal supersaturated
\vapour{}, has its limits: if nucleation occurs too fast, it affects the properties
of the \vapour{} to a significant extent so that the nucleation rate obtained according
to the method of Yasuoka and Matsumoto \cite{YM98} and other
properties of the system vary over simulation time \cite{CWSWR09}.
In the present work, nucleation is studied as a steady-state process by
combining grand canonical MD (GCMD) and McDonald's \daemon{} \cite{McDonald62, HV09},
an \qq{intelligent being} that eliminates large droplets from the system.

\section{Simulation method}
Supersaturated states can be characterized in terms of 
the difference between the chemical potential $\chempot$
of the \vapour{} and the saturated chemical potential $\sat{\chempot}(\temperature)$.
The chemical potential of the \vapour{} can be regulated
by simulating the grand canonical ensemble with GCMD: alternating
with canonical ensemble MD steps, particles are inserted into and deleted from
the system probabilistically, with the usual grand canonical acceptance criterion \cite{Cielinski85}.
For a test insertion, random coordinates are chosen for an additional particle,
and for a test deletion, a random particle is removed from the system.
The potential energy difference $\Delta\upot$ due to the test action is determined
and compared with the chemical potential.
The acceptance probability for insertions is
\begin{equation}
   \probability = \min\left(1,
      \exp\left[\frac{\chempot - \Delta\upot}{\kboltz\temperature}\right]
         \frac{\volume}{\thermal^3(\absnum+1)} \right),
\end{equation}
while for deletions it is
\begin{equation}
   \probability = \min\left(1,
      \exp\left[\frac{-\chempot - \Delta\upot}{\kboltz\temperature}\right]
         \frac{\volume}{\thermal^3\absnum} \right),
\end{equation}
wherein $\thermal$ is the thermal wavelength. Of course, care must be
taken that the momentum of the inserted particles is consistent with
the simulated ensemble and does not introduce any artifical velocity gradients.
The MD integration time step was $\Delta\timea$ = 0.00404
in reduced time units,
i.e., $\LJsig(\mass\slash\LJeps)^{1/2}$, wherein 
$\LJeps$ is the energy parameter of the fluid model and
$\mass$  is the mass of a particle.
The number of test actions per simulation time step
was chosen between $10^{-6}$ and $10^{-3}$ $\absnum$, a value which was occasionally
decreased after equilibration if very low nucleation rates were observed.

Molecular simulation of nucleation has to rely on a cluster criterion to distinguish
the emerging liquid from the surrounding supersaturated \vapour{} \cite{BM01}.
In the present case, the Stillinger criterion \cite{Stillinger63} was
used to define the liquid phase and \nuclei{} were determined as biconnected
components. Whenever a \nucleus{} exceeded the specified
threshold size $\threshold$, an intervention
of McDonald's \daemon{} removed it from the system, leaving a
vacuum behind \cite{McDonald62, HV09}.

\section{Nucleation theory}

\noindent
The free energy of \nucleus{} formation is the same for the grand canonical
and the isothermal-isobaric ensemble \cite{ZRR06}. At specified values of
the chemical potential $\chempot$ of the supersaturated \vapour{},
the total system volume $\volume$ and the temperature $\temperature$, it is related to
the surface energy $\surfaceenergy$ by \cite{Debenedetti96}
\begin{equation}
   \Deltagrandpotentialof{\nuclsize}
      = \int_{\nucvolume(1)}^{\nucvolume(\nuclsize)}
         (\pressure - \nucpressure) \differential\nucvolume
            + \int_{\areaof{1}}^{\areaof{\nuclsize}}
               \left(\frac{\partial\surfaceenergy}{\partial\area}\right)
                  \differential\area
                     + \int_{1}^{\nuclsize} (\chempotof{\nuclsize} - \chempot)
	                \differential\nuclsize,
\end{equation}
where $\nuclsize$ is the number of particles in the \nucleus{},
$\pressure$ is the supersaturated \vapour{} pressure,
$\nucvolume(\nuclsize)$ is the volume and $\areaof{\nuclsize}$ the surface
area of a \nucleus{} containing $\nuclsize$ particles. Note that
$\chempotof{\nuclsize}$ as well as $\nucpressure$ are the chemical
potential and the pressure of the liquid phase
at the conditions prevailing inside the \nucleus{}.
%
In CNT, it is assumed that the bulk liquid density at saturation $\liq{\density}$ and the
density of a nanoscopic \nucleus{} are the same and all
\nuclei{} are treated as spheres, i.e., $\liqdensity = \liq{\density}$ 
and $\areaof{\nuclsize} = \areasphof{\nuclsize} = 
\left(6\sqrt{\pi}\nuclsize\slash\liq{\density}\right)^{2\slash{}3}$.
Accordingly, the chemical potential of the liquid
inside the nucleus is approximated by
\begin{equation}
   \chempotof{\nuclsize} = \sat\chempot(\temperature) +
      \int_{\sat\pressure}^{\nucpressure} 
         \frac{\differential\pressure}{\liqdensity}
	    \approx \sat\chempot(\temperature)
	       + \frac{\nucpressure - \sat\pressure(\temperature)}{\liq\density},
\end{equation}
and the \nucleus{} surface tension $\curvedtension = (\partial\surfaceenergy\slash\partial\area)$
by the surface tension $\planartension$ of the planar \vapour-liquid interface,
leading to \cite{RBDR03, WHBR08}
\begin{equation}
   \differential\grandpotential = \left[
      \planartension \, \sqrt[3]{\frac{2\pi}{3\nuclsize}\left(\frac{4}{\liq\density}\right)^2}
         + \sat\chempot(\temperature) - \chempot
	    + \frac{\pressure - \sat\pressure(\temperature)}{\liq\density}
               \right] \differential\nuclsize.
\end{equation}
The free energy of formation has a
maximum $\crit{\Delta\grandpotential}$ which lies at the size $\crit{\nuclsize}$
of the critical nucleus.
Including the Zel'dovi\v{c}
factor $\zeldovich$ and the 
thermal non-accomodation factor $\noniso$ of Feder \textit{et al.\ }\cite{FRLP66}, the
nucleation rate is
\begin{equation}
   \nuclrate = \noniso\zeldovich \frac{\absnumof{1}}{\volume}
      \exp(-\invkT\crit{\Delta\grandpotential})
         \frac{\pressure\thermal}{\planck}\areaof{\crit{\nuclsize}},
\end{equation}
where $\absnumof{1}$ is the number of \vapour{} molecules in the system
and $\planck$ is the Planck constant.

%
Instead of using the surface tension of the planar interface, Laaksonen, Ford, and
Kulmala (LFK) \cite{LFK94} proposed an expression equivalent to
\begin{equation}
   \int_{0}^{\areaof{\nuclsize}} \curvedtension\differential\area
      = \planartension\areaof{\nuclsize}\left( 1
         + \LFKa\nuclsize^{-1\slash{}3} + \LFKb\nuclsize^{-2\slash{}3}
            \right).
\end{equation}
The two parameters $\LFKa$ and $\LFKb$ are determined from the assumption that almost all particles
are arranged either as monomers or as dimers and that the
Fisher \cite{Fisher67} equation of state correctly relates $\pressure\slash\temperature$ to the
number of monomers and \nuclei{} present per volume. Effectively, LFK theory modifies CNT only
by the introduction of the parameter $\LFKa$, since $\LFKb$
cancels out for all free energy differences if the usual assumption $\area \sim \nuclsize^{2\slash{}3}$
is applied.

%
The Hale scaling law (HSL) is based on a different approach \cite{Hale86}. 
In agreement with experimental data on nucleation of water and toluene \cite{Hale86}, it predicts
\begin{equation}
   \nuclrate \sim \density^{-2\slash{}3}\left(\frac{\planartension}{\temperature}\right)^{1\slash{}2}\pressure^2
      \exp\left[ \frac{4\planartension^3}{27 (\ln\supersat)^2} \right],
\end{equation}
with a proportionality constant depending only on properties of the critical point.

\begin{figure}[b]
\centering
\includegraphics[width=10cm]{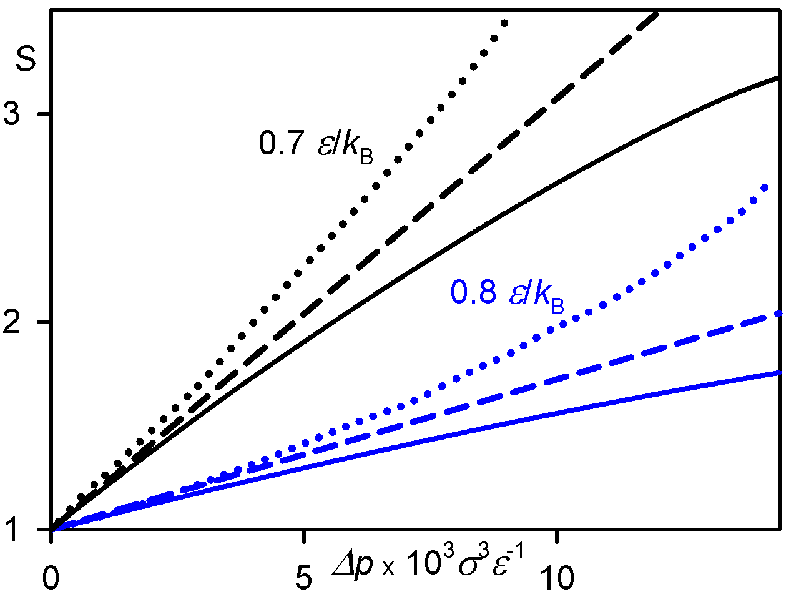}
\caption{Chemical potential supersaturation $\supersat$ (---),
pressure supersaturation $\pressure\slash\sat{\pressure}$ $\textnormal{(-- --)}$,
and density supersaturation $\density\slash\sat{\density}$
$\textnormal{($\cdot$ $\cdot$ $\cdot$)}$
in dependence of the excess pressure
$\Delta\pressure = \pressure - \sat{\pressure}$ at
$\temperature$ = 0.7 and 0.8 $\LJeps\slash\kboltz$.}
\label{szilI}
\end{figure}

\begin{figure}[t]
\centering
\includegraphics[width=10cm]{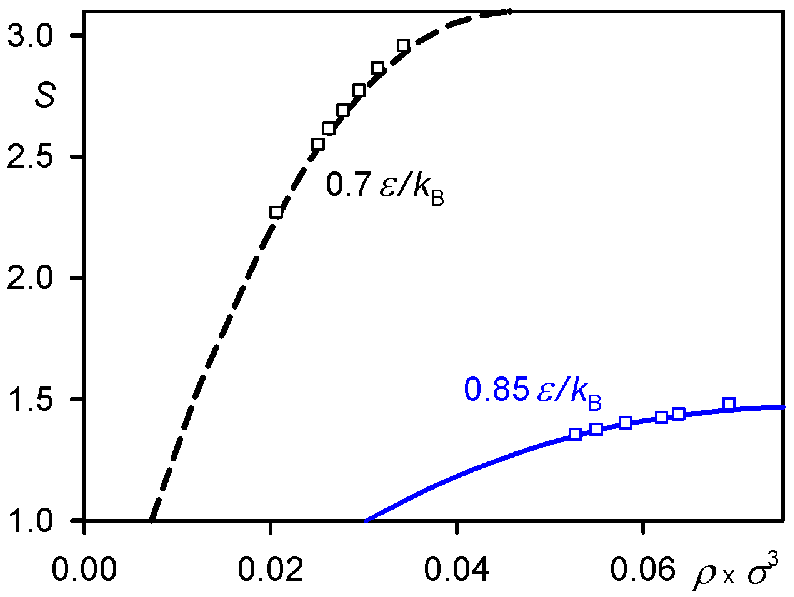}
\caption{Density dependence of the chemical potential supersaturation for the \vapour{} of the \LJTS{} fluid,
obtained from GCMD simulation with McDonald's \daemon{} ($\square$) and by
integration of the Gibbs-Duhem equation using data from canonical ensemble MD
simulation with $\temperature$ = 0.7 $\textnormal{(-- --)}$
and 0.85 $\LJeps\slash\kboltz$ (---).}
\label{szilIII}
\end{figure}

%
In the present work, these theories are evaluated using Gibbs-Duhem integration
over the metastable part of the \vapour{} pressure isotherm
collected by canonical ensemble MD simulation of small systems.
The fluid model under consideration is the truncated and shifted Lennard-Jones (\LJTS{}) potential
with a cutoff radius of $2.5 \LJsig$ \cite{AT87}.
Note that the chemical potential supersaturation, i.e.,
$\supersat = \exp\left(\invkT[\chempot - \sat\chempot(\temperature)]\right)$,
deviates considerably from the pressure
supersaturation $\pressure\slash\sat{\pressure}$ and the
density supersaturation $\density\slash\sat{\density}$, with respect to the saturated
\vapour{} pressure $\sat{\pressure}(\temperature)$ and density $\vap{\density}(\temperature)$
of the bulk, cf.\ Fig.\ \ref{szilI}.
For the saturated chemical potential of the \LJTS{} fluid, a correlation based on
previously published data \cite{VKFH06} gives
\begin{equation}
   \frac{\sat{\chempot}(\temperature) - \chempotid(\temperature)}{\kboltz\temperature} = 
      -0.2367  - \frac{1.7106 \LJeps}{\kboltz\temperature} - \frac{1.1514 \LJeps^2}{(\kboltz\temperature)^2}.
\end{equation}
In Fig.\ \ref{szilIII}, the chemical potential supersaturation is shown as a function of the
\vapour{} density determined by GCMD simulation with McDonald's \daemon{}.
These values agree well with the metastable \vapour{} pressure isotherm of the
\LJTS{} fluid obtained by canonical ensemble simulation. 

\section{Intervention rate and nucleation rate}

The size evolution of any given \nucleus{} can
be considered as a random walk over the order parameter $\nuclsize$, changing
only by relatively small amounts $\Delta\nuclsize$, usually by
the absorption or emission of monomers.
As discussed by Smoluchowski \cite{Smoluchowski08, Smoluchowski15} during
his scientifically most productive period in L'viv and Krak\'ow, the probabilities for
the growth and decay transitions are proportional to the respective values of
the partition function $\partition$, resulting in
\begin{equation}
   \probgrowof{\nuclsize} = \frac{1}{2}
       + \frac{(\differential\partition\slash\differential\nuclsize) \Delta\nuclsize
          }{2\partition + \landau(\nuclsize^2)} + \landau(\nuclsize^2),
\end{equation}
and
\begin{equation}
   \probshrinkof{\nuclsize} = \frac{1}{2}
       - \frac{(\differential\partition\slash\differential\nuclsize) \Delta\nuclsize
          }{2\partition + \landau(\nuclsize^2)} + \landau(\nuclsize^2).
\end{equation}
The probability $\probfinallyof{\nuclsize}$
that a certain size is \textit{eventually} reached (at any time during the
random walk process), given that the current size is $\nuclsize$, has the property
\begin{equation}
   \probfinallyof{\nuclsize}
      = \probgrowof{\nuclsize}\probfinallyof{\nuclsize + \Delta\nuclsize}
         + \probshrinkof{\nuclsize}\probfinallyof{\nuclsize - \Delta\nuclsize}.
\end{equation}
By substituting
\begin{equation}
   \probfinallyof{\nuclsize \pm \Delta\nuclsize}
      = \probfinallyof{\nuclsize}
         \pm \frac{\differential\probfinally}{\differential\nuclsize} \Delta\nuclsize
	    + \frac{\differential^2\probfinally}{2\differential\nuclsize^2}
	       \Delta\nuclsize^2 + \landau\left(\Delta\nuclsize^3\right),
\end{equation}
it follows for small $\Delta\nuclsize$ neglecting terms of third order and beyond, that
\begin{equation}
   \frac{\differential\partition}{\partition\differential\nuclsize}
      = \frac{-\differential\left(\differential\probfinally\slash\differential\nuclsize\right)
         }{2\left(\differential\probfinally\slash\differential\nuclsize\right)
	    \differential\nuclsize}.
\end{equation}
Using the partition function for the grand canonical ensemble, the
derivative of the probability is given by
\begin{equation}
   \frac{\differential\probfinally}{\differential\nuclsize}
      = \pconstant \exp\left(2 \invkT \Deltagrandpotentialof{\nuclsize}\right),
\end{equation}
where $\pconstant$ is an integration constant.
Obtaining the two remaining parameters from the boundary conditions
\begin{eqnarray}
   \probinftyof{1} &=& 0, \\
   \lim_{\threshold\to\infty} \probinftyof{\threshold} &=& 1,
\end{eqnarray}
the probability $\probinftyof{\threshold}$ for
a \nucleus{} containing $\threshold$ molecules of eventually reaching macroscopic size,
i.e., $\nuclrate\to\infty$, is
\begin{equation}
   \probinftyof{\threshold}
      = \frac{\int_1^\threshold
         \exp\left(2 \invkT \Deltagrandpotentialof{\nuclsize}\right) \differential\nuclsize
            }{ \int_1^\infty
	       \exp\left(2 \invkT \Deltagrandpotentialof{\nuclsize}\right) \differential\nuclsize
	       }.
   \label{eqn:intervention}
\end{equation}
The intervention rate $\nuclrateof{\threshold}$ of McDonald's \daemon{} is
related to the nucleation rate $\nuclrate$ by
\begin{equation}
   \nuclrate = \nuclrateof{\threshold}\probinftyof{\threshold}.
   \label{eqn:intervention2}
\end{equation}
Thus, with an intervention threshold far below the critical size, the
intervention rate is many orders of magnitude
higher the steady-state nucleation rate. However, as confirmed by the present simulation
results shown in Tab.\ \ref{tabtheta}, it reaches a plateau
for $\threshold > \crit{\nuclsize}$, where $\crit{\nuclsize} = 41$ according to
CNT and 39 according to SPC.

\begin{table}[b]
\caption{
   Dependence of the intervention rate $\nuclrateof{\threshold}$ as well as the
   probability $\probinftyof{\threshold}$ according to CNT and LFK
   on the intervention threshold size $\threshold$ for
   McDonald's \daemon{} during GCMD simulation at
   $\temperature = 0.7$ $\LJeps\slash\kboltz$ and $\supersat = 2.4958$, where
   the rates are given in units of $(\LJeps\slash\mass)^{1\slash{}2}\LJsig^{-4}$.
   The number of particles in the system
   and the values for the
   pressure supersaturation $\pressure\slash\sat{\pressure}$
   refer to the steady state and the constant
   volume of the system is given in units
   of $\LJsig^3$.
   \label{tabtheta}
}
\begin{tabular}{ccc|ccc|cc}
   $\volume$ & $\absnum$ & $\pressure\slash\sat\pressure$
             & $\threshold$
	     & $\ln\probinftyof{\threshold}(\mathrm{CNT})$
	     & $\ln\probinftyof{\threshold}(\mathrm{LFK})$
	     & $\ln\nuclrateof{\threshold}$ \\ \hline
      5.38 $\times$ 10$^6$ & \phantom{0}124000 & 2.70
      & 10 & -16.7\phantom{00} & -12.7\phantom{00}
      & -13.6 \\
      4.32 $\times$ 10$^7$ & 1020000 & 2.75
      & 20 & \phantom{0}-8.14\phantom{0} & \phantom{0}-6.33\phantom{0}
      & -17.0 \\
      5.38 $\times$ 10$^6$ & \phantom{0}129000 & 2.78
      & 25 & \phantom{0}-5.55\phantom{0} & \phantom{0}-4.34\phantom{0}
      & -17.6 \\
      5.38 $\times$ 10$^6$ & \phantom{0}129000 & 2.78
      & 35 & \phantom{0}-2.32\phantom{0} & \phantom{0}-1.82\phantom{0}
      & -19.9 \\
      4.32 $\times$ 10$^7$ & 1040000 & 2.78
      & 48 & \phantom{0}-0.508 & \phantom{0}-0.400
      & -21.7 \\
      4.32 $\times$ 10$^7$ & 1040000 & 2.78
      & 65 & \phantom{0}-0.022 & \phantom{0}-0.019
      & -21.9 \\
      2.15 $\times$ 10$^7$ & \phantom{0}518000 & 2.77
      & 74 & \phantom{0}-0.002 & \phantom{0}-0.002
      & -22.1
\end{tabular}
\end{table}

\section{Results and discussion}

Homogeneous nucleation of the \LJTS{} fluid was studied by a series of GCMD simulations
with McDonald's \daemon{} for systems containing up to 17 million particles.
%

\begin{figure}[t!]
\centering
\includegraphics[width=10cm]{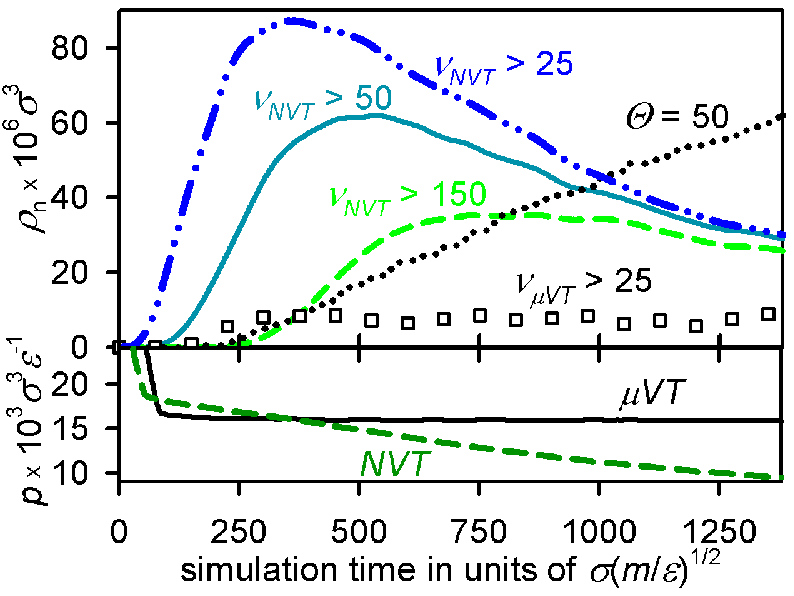}
\caption{
   Top: Number per unit volume $\nucdensity$ of \nuclei{} containing more than
        25 $\textnormal{($\cdot$ -- $\cdot$)}$, 50 (---),
        and $\textnormal{150 (-- --)}$ particles
        in a canonical ensemble MD simulation at
        $\temperature$ = 0.7 $\LJeps\slash\kboltz$ and $\density$ = 0.004044
        $\LJsig^{-3}$ using a hybrid geometric-energetic cluster criterion,
        number per unit volume $\nucdensity$ of \nuclei{} with $\nuclsize \geq 25$ ($\square$)
        in a GCMD simulation with $\temperature$ = 0.7 $\LJeps\slash\kboltz$, $\supersat$ = 2.8658,
        and $\threshold$ = 50, using the Stillinger \cite{Stillinger63} cluster
        criterion with \nuclei{} determined as biconnected components,
        as well as the aggregated number of McDonald's \daemon{} interventions per unit volume
        in the GCMD simulation, over simulation time.
   Bottom: Pressure over simulation time for the canonical ensemble
           MD simulation $\textnormal{(-- --)}$
           and the GCMD simulation with McDonald's \daemon{} (---) \cite{HV09}.
}
\label{szilII}
\end{figure}

%
%
After a temporal delay, depending on the threshold size, the
pressure and the intervention rate reached a constant value, cf.\
Fig.\ \ref{szilII}.
In a canonical ensemble MD simulation under similar conditions as
the GCMD simulation that is also shown in Fig.\ \ref{szilII}, the pressure supersaturation
decreased from about 3 to 1.5 and the rate of formation was significantly lower for
larger nuclei, due to the free energy effect accounted for by
Eqs.\ (\ref{eqn:intervention}) and (\ref{eqn:intervention2})
as well as the depletion of the \vapour{} \cite{HV09}.

The constant
supersaturation of the GCMD simulation agreed approximately with the
time-dependent supersaturation in the canonical ensemble about
$\timea$ = 400 after simulation onset, cf.\ Fig.\ \ref{szilII}.
At this stage, the number of
small \nuclei{} present per volume was similar in both cases, and
the rate of formation for \nuclei{} with $\nuclsize > 150$ at $\timea$ = 400
in the canonical ensemble simulation was of the same order of magnitude as the
intervention rate of the \daemon{}.

Van Meel \textit{et al.}\ \cite{MPSF08} determined by MC simulation with
forward flux sampling that supersaturated \vapours{} of the \LJTS{} fluid
at a temperature of $\temperature$ = 0.45 $\LJeps\slash\kboltz$, i.e.,
significantly below the triple point $\Ttr$ = 0.65 $\LJeps\slash\kboltz$,
initially undergo \vapour{} to liquid nucleation, and
CNT is known to underestimate the \vapour{} to liquid nucleation rate of unpolar
fluids \cite{HVH08}.
The present \daemon{} intervention rates confirm this conclusion.
LFK and HSL are significantly more accurate than CNT.
Note that in Tab.\ \ref{tab045}, the nucleation rate according to
Eq.\ (\ref{eqn:intervention2}) based on the CNT value of $\probinftyof{\threshold}$
is given.

From Tab.\ \ref{tab045} it is also confirmed that the \qq{direct observation
method} (DOM) \cite{CWSWR09}, which in the present case corresponds to
assuming
\begin{equation}
   \ln\nuclrateof{\threshold} = \ln\nuclrate - \ln\probinftyof{\threshold}
      = -\ln\delay\volume,
\end{equation}
where $\delay$ is the temporal delay of formation for the first sufficiently large
\nucleus, is inadequate for nucleation near the spinodal line.

\begin{table}[b!]
\caption{
   \Vapour{} to liquid nucleation rate at $\temperature$ = 0.45 $\LJeps\slash\kboltz$ from
   GCMD simulation with McDonald's \daemon{}.
   The theories were evaluated with respect to the metastable \vapour-liquid
   equilibrium at $\sat{\pressure} = 4.28 \times 10^{-5}$ $\LJeps\slash\LJsig^3$ \cite{MPSF08},
   and the \vapour-liquid surface tension
   $\planartension = 1.07$ $\LJeps\slash\LJsig^2$ \cite{MPSF08}
   was used. 
   \label{tab045}
}
\begin{tabular}{cc|ccccc|ccc}
   $\pressure\slash\sat{\pressure}$ & $10^{-6}\absnum$
      & $\threshold$ & $-\ln\delay\volume$
         & $\ln\probinftyof{\threshold}(\mathrm{CNT})$ & $\ln\nuclratesim$ &
            & $\ln\nuclrateCNT$ & $\ln\nuclrateLFK$ & $\ln\nuclrateHale$
	       \\ \hline
   30.2\phantom{0} & \phantom{0}0.397
      & \phantom{0}9 & -23.1
         & -4.57\phantom{0} & \phantom{0}-26.4 &
            & \phantom{0}-31.5 & \phantom{0}-26.2 & \phantom{0}-24.7 \\
   32.4\phantom{0} & \phantom{0}0.429
      & \phantom{0}9 & -23.0
         & -3.80\phantom{0} & \phantom{0}-25.0 &
            & \phantom{0}-30.5 & \phantom{0}-25.4 & \phantom{0}-24.0 \\
   55.9\phantom{0} & \phantom{0}1.07\phantom{0}
      & 12 & -22.5
         & -0.062 & \phantom{0}-18.0 &
            & \phantom{0}-24.2 & \phantom{0}-20.2 & \phantom{0}-19.5 \\
   74.7\phantom{0} & 17.1\phantom{00}
      & 24 & -17.1
         & $\approx$ 0 & \phantom{0}-18.8 &
            & \phantom{0}-21.8 & \phantom{0}-18.6 & \phantom{0}-17.7
\end{tabular}
\end{table}

\section{Conclusion}
GCMD with McDonald's \daemon{} was established as a method for steady-state simulation of
nucleating \vapours{} at high supersaturations. 
A series of simulations was conducted for the \LJTS{} fluid.
CNT was found to underpredict the nucleation rate below the triple
point, whereas LFK and HSL more accurately describe \vapour{} to liquid
nucleation of the \LJTS{} fluid.

\begin{theacknowledgments}
   The authors would like to thank G.\ Chkonia, H.\ Hasse, S.\ Sastry,
   C.\ Valeriani, and J.\ Wedekind for fruitful discussions
   and Deutsche Forschungsgemeinschaft for funding SFB 716. The presented research
   was conducted under the auspices of the Boltzmann-Zuse Society of
   Computational Molecular Engineering (BZS), and the simulations were
   performed on the HP XC4000 supercomputer at
   the Steinbuch Centre for Computing, Karlsruhe, under the grant LAMO,
   as well as the \textit{phoenix} supercomputer at H\"ochstleistungs\-rechenzentrum
   Stuttgart (HLRS) under the grant MMHBF.
\end{theacknowledgments}


\end{document}